\documentclass[
  ,final            % use final for the camera ready runs
  ]
  {aipproc}

\layoutstyle{6x9}

\newcommand{\rmd}{\ensuremath{{\rm d}}}

\begin{document}

\title{Hawking radiation as perceived by different observers}


\author{Luis~C.~Barbado}{
  address={Instituto de Astrof\'{\i}sica de Andaluc\'{\i}a (CSIC), Glorieta de la Astronom\'{\i}a, 18008 Granada, Spain}
}

\author{Carlos~Barcel\'o}{
  address={Instituto de Astrof\'{\i}sica de Andaluc\'{\i}a (CSIC), Glorieta de la Astronom\'{\i}a, 18008 Granada, Spain}
}

\author{Luis~J.~Garay}{
  address={Departamento de F\'{\i}sica Te\'orica II, Universidad Complutense de Madrid, 28040 Madrid, Spain}
  ,altaddress={Instituto de Estructura de la Materia (CSIC), Serrano 121, 28006 Madrid, Spain}
}

\begin{abstract}

We study the perception of Hawking radiation by different observers outside a black hole. The analysis is done in terms of an effective-temperature function that varies along the trajectory of each observer. The vacuum state of the radiation field is chosen to be non-stationary, so as to mimic the switching-on of Hawking radiation that would appear in a real black hole collapse. We analyse how this vacuum is perceived by observers staying at a fixed radius, by observers coming in free-fall from radial infinity at different times, and by observers in free-fall released from finite radial positions. Results found have a compelling physical interpretation. One main result, at first unexpected, is that in general free-falling observers do perceive particle emission by the black hole when crossing the event horizon. This happens because of a diverging Doppler shift at the event horizon.

\end{abstract}

\maketitle

%%%%%%%%%%%%%%%%%%%%%%%%%%%%%%%%%%%%%%%%%%%%
%% MAINMATTER
%%%%%%%%%%%%%%%%%%%%%%%%%%%%%%%%%%%%%%%%%%%%

\section{Introduction}

The discovery by Hawking that black holes are not as black as they appear in classical general relativity, but that they emit a flux of particles with thermal spectrum~\cite{Hawking:1974sw}, the so called \emph{Hawking radiation,} is arguably the most important result in the study of quantum fields in curved spacetime. The temperature of the radiation emitted is proportional to the surface gravity of the black hole, and thus inversely proportional to its mass. Later on, Unruh pointed that, for the black hole to emit this thermal radiation, the quantum field should be in a particular vacuum state, called the \emph{Unruh vacuum state}~\cite{Unruh:1976db}. This vacuum state corresponds to the mode decomposition associated to observers in free-fall and arbitrarily close to the event horizon. Thus, as it was already well known, perception of the quantum radiation in a curved geometry was an observer dependent notion.

Since the discovery of this radiation phenomenon, many different procedures have been applied to study its nature and characteristics. In our work~\cite{Barbado:2011dx}, we analyse the perception of Hawking radiation by observers following different trajectories outside a Schwarzschild black hole. This method is based on the calculation of an \emph{effective-temperature function} introduced in~\cite{Barcelo:2010pj} that varies with time, and that gives different results for the different trajectories of the observers. Under certain conditions for this function, one can assure that the observer is perceiving thermal radiation from the field, and that its temperature is proportional to the value of this effective-temperature function.

Although we work in a $3+1$ Schwarzschild geometry, we restrict ourselves to the radial sector. Also, we consider for simplicity a massless Klein-Gordon scalar field as the radiation field. The usual choice for the quantum state of the field is the Unruh vacuum state~\cite{Unruh:1976db}. As we already mentioned, in this state the black hole emits thermal Hawking radiation at radial infinity, with temperature proportional to the surface gravity of the black hole. Instead of this vacuum, we choose a different, non-stationary vacuum state, that interpolates smoothly in time between the Boulware vacuum state~\cite{Boulware:1974dm} (with no radiation emission) and the Unruh vacuum state. In this way, we mimic the switching-on process of Hawking radiation that would occur in a realistic collapse scenario (with a non-static geometry).

Here, we will show the results found for two kinds of observers: static observers at a fixed radius, and observers left to free-fall from infinity at different times. In~\cite{Barbado:2011dx}, we deal with a third case of observers: those left to free-fall from a finite radius. Some general discussion on these results is also given.

\subsection{Effective-temperature function}

As we already said, our work is based on an effective-temperature function introduced in~\cite{Barcelo:2010pj}. This function is defined along the trajectory of any observer as a function of its proper time. It depends on the particular trajectory of the observer, but also on the vacuum state choice. The selection of the vacuum state can be encoded in the definition of an outgoing null coordinate $U$ (see~\cite{Barbado:2011dx, Barcelo:2010pj} for details). Once a vacuum state has been chosen, we consider the timelike trajectory of an specific observer, and introduce its proper time $u$ as a new outgoing null coordinate (this observer can `label' the outgoing null rays it crosses by using its proper time). With both null coordinates properly defined, we can construct the relation $U(u)$. The effective-temperature function is defined as
\begin{equation}
\kappa(u) := -\left. \frac{\rmd^2 U}{\rmd u^2} \middle/ \frac{\rmd U}{\rmd u}. \right. \nonumber
\end{equation}

As we can see, it is a function of the proper time $u$ of the observer. If the variation of this function is slow enough, we say that the \emph{adiabatic condition} is satisfied, and one can assure that the observer is perceiving a Plankian spectrum of particles with temperature $T = \kappa/(2 \pi k_{\rm B})$, proportional to $\kappa$ (in the following, we omit mentioning the proportionality factor). This adiabatic condition is defined by
\begin{equation}
\varepsilon(u) := \frac{1}{\kappa^2}\left| \frac{\rmd \kappa}{\rmd u} \right| \ll 1. \nonumber
\end{equation}
In the case that this is not satisfied, we cannot talk about thermal perception, although we will still consider $\kappa (u)$ as an estimator of the amount of particles perceived at each instant of the trajectory.

In particular, in our work we choose the null coordinate $U$ as the proper time (via ``ray labelling'') of one particular fixed observer left in free-fall from the radial infinity. We shall call this vacuum state the \emph{collapse vacuum.} It is non-stationary, and it interpolates smoothly between the Boulware and the Unruh vacuum states.

\section{Results for different observers}

\subsection{Observers static at a fixed radius}

\begin{figure}[ht]
\begin{minipage}[b]{0.475\linewidth}
\centering
\includegraphics[width=\linewidth]{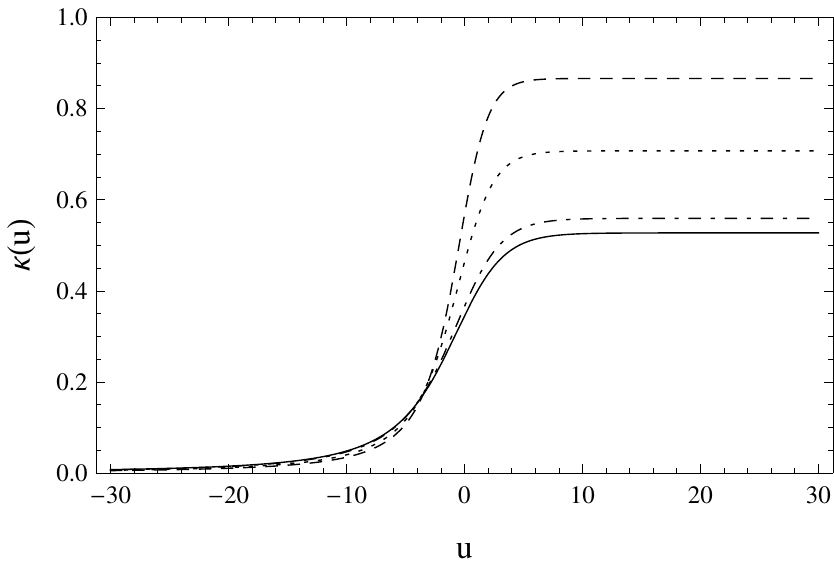}
\end{minipage}
\hspace{0.04\linewidth}
\begin{minipage}[b]{0.475\linewidth}
\centering
\includegraphics[width=\linewidth]{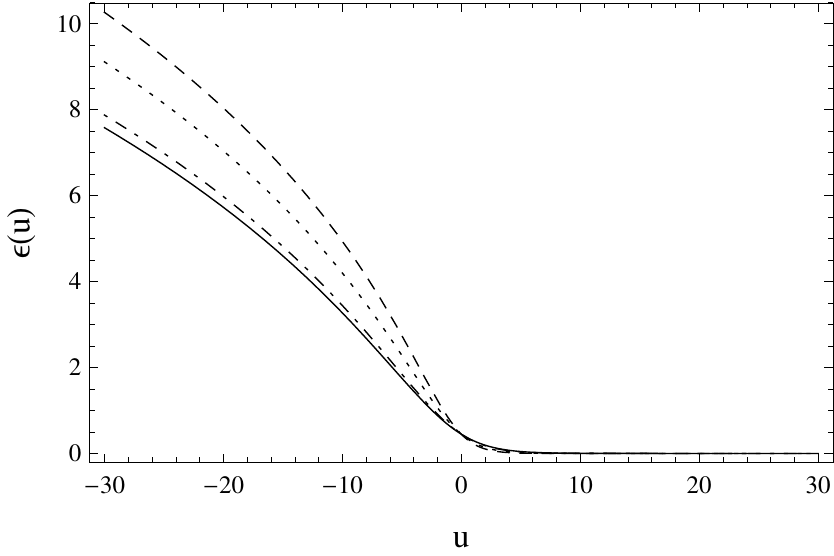}
\end{minipage}
\caption{$\kappa(u)$ and $\varepsilon(u)$ for different static observers at $r_0 = $ [$3 m$ (dashed), $4 m$ (dotted), $10 m$ (dash-dot), $20 m$ (solid)]. We use $2m=1$ units. (Figures reprinted from~\cite{Barbado:2011dx}.)}
\label{static}
\end{figure}

The effective-temperature function for observers staying at a fixed radius $r_0$ is plotted in figure~\ref{static}, together with the parameter $\varepsilon(u)$ that controls the adiabatic condition.

\noindent

We can check that the switching-on process of the radiation is verified. Observers initially detect no radiation, and, after a transient period, they perceive almost perfect thermal radiation (the value of $\varepsilon(u)$ crushes to nearly zero). Note that the huge value of $\varepsilon(u)$ at earlier times has no real meaning, as in that region the value of $\kappa(u)$ is nearly zero, and therefore there is virtually no radiation. The temperature of the final radiation for observers far enough from the horizon is $\kappa \simeq 1/(4m)$, which is the Hawking temperature. Observers closer to the horizon experiment a gravitational blue-shift.

\subsection{Observers free falling from infinity}

\begin{figure}[ht]
\begin{minipage}[b]{0.475\linewidth}
\centering
\includegraphics[width=\linewidth]{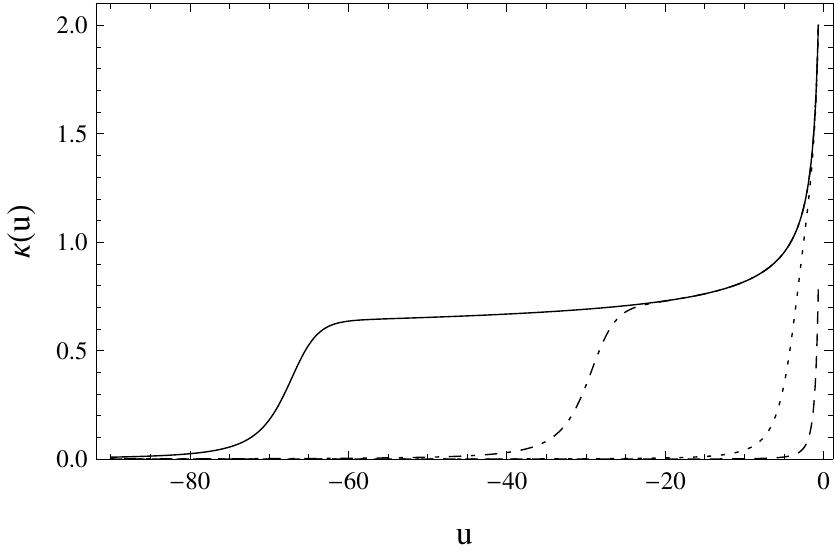}
\end{minipage}
\hspace{0.04\linewidth}
\begin{minipage}[b]{0.475\linewidth}
\centering
\includegraphics[width=\linewidth]{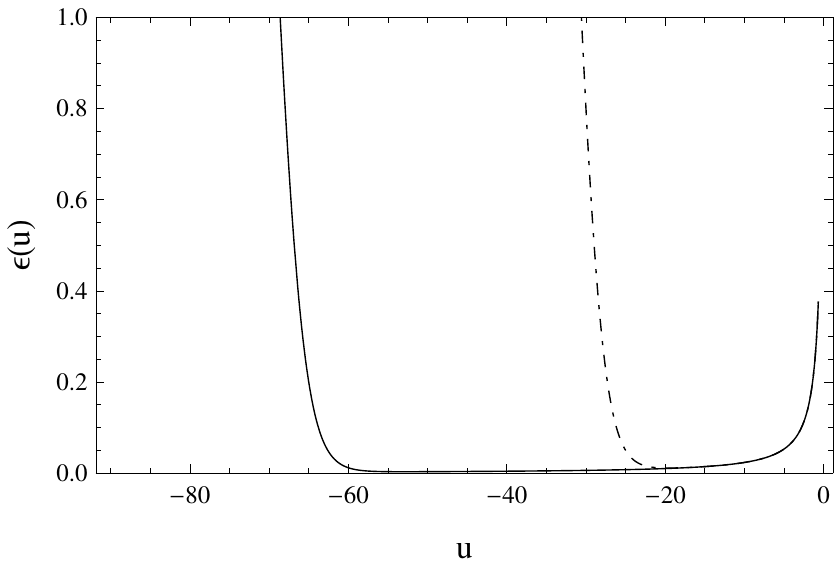}
\end{minipage}
\caption{$\kappa(u)$ and $\varepsilon(u)$ for different free-falling observers with $\Delta t_0 = $ [$2 m$ (dashed), $20 m$ (dotted), $100 m$ (dash-dot), $200 m$ (solid)]. We use $2m=1$ units. (Figures reprinted from~\cite{Barbado:2011dx}.)}
\label{falling}
\end{figure}

These observers follow the same kind of trajectory as the observer defining the vacuum state. However, they will have a finite delay in time $\Delta t_0$ with respect to that one. In figure~\ref{falling}, we plot the effective-temperature function and the parameter $\varepsilon(u)$ for observers with different delays. All of them finally cross the event horizon at some proper time, that we have arbitrarily chosen to be $u=0$.

We can see here again the switching-on of Hawking radiation. After the transient period, observers with enough delay $\Delta t_0$ reach a `plateau' in the effective-temperature function. The variation of the function along this period is slow (again, $\varepsilon(u) \simeq 0$), and thus the radiation perceived is thermal to a good approximation. Its value is the Hawking temperature, but augmented because of the gravitational and Doppler blue-shifts.

Finally, observers detect a peak in the effective-temperature before crossing the event horizon. At first sight, this is contradictory with the fact that the Unruh vacuum state is associated to the mode decomposition that observers in free-fall and close to the horizon would naturally do~\cite{Unruh:1976db}. Let us remind that, for times late enough, the non-stationary state we are considering is nearly identical to the Unruh state. Of course, this paradox is solvable. In fact, it can be shown~\cite{Barbado:2011dx, Barbado:2012} that there is a free-falling observer arbitrarily close to the horizon that detects no radiation (even when the quantum field is radiative), namely the one which is also at rest with respect to the black hole. Observers with non-zero radial velocity towards the black hole suffer a diverging Doppler blue-shift with respect to this one. This divergence ``competes'' with the vanishing of the effective-temperature for the instantaneously static observer, and the result happens to be a finite value for $\kappa$. In the case in which the field is already in the Unruh vacuum state, it can be shown that, for observers in free-fall from the radial infinity, this value is $\kappa = 1/m$ (four times the Hawking temperature). This is also the result for the observers with enough delay in figure~\ref{falling}. One can see that the adiabatic condition is not strictly satisfied at the final peak ($\varepsilon(u \to 0) \simeq 3/8$), but we are not too far from the adiabatic regime either. The radiation spectrum perceived there will not differ strongly from a Planckian shape.

\section{Further work}

We can conclude that results found when calculating the effective-temperature function for different observers are compelling, and can be physically interpreted in terms of well-known phenomena. In a more recent work~\cite{Barbado:2012}, the authors find a general expression for the effective-temperature function that depends only on the vacuum state choice and on the local properties of the observer's trajectory (position, velocity and acceleration). At present, this formula is being exploited for analysing buoyancy effects in black holes.

\bibliographystyle{aipproc}
\bibliography{proceeding}

\end{document}